\documentstyle[11pt]{article}
\begin{document}
\begin{titlepage}
\begin{center}
{\Large \bf Soft SUSY Breaking Parameters and RG Running of Squark and Slepton Masses in Large Volume Swiss Cheese Compactifications}
\vskip 0.1in { Aalok Misra\footnote{e-mail: aalokfph@iitr.ernet.in} and
Pramod Shukla \footnote{email: pmathdph@iitr.ernet.in}\\
Department of Physics, Indian Institute of Technology,
Roorkee - 247 667, Uttaranchal, India}
\vskip 0.5 true in
\end{center}
\thispagestyle{empty}

\begin{abstract}
  We consider type IIB large volume compactifications involving orientifolds of the Swiss-Cheese Calabi-Yau ${\bf WCP}^4[1,1,1,6,9]$
 with a single mobile space-time filling $D3$-brane  and stacks of $D7$-branes wrapping the
 ``big" divisor $\Sigma_B$ (as opposed to the ``small" divisor usually done in the literature thus far) as well as supporting $D7$-brane fluxes.
  After reviewing our proposal of \cite{susyrevelation} for resolving a long-standing tension between large volume cosmology and phenomenology pertaining to obtaining a $10^{12} GeV$ gravitino in the inflationary era and a $TeV$ gravitino in the present era, and summarizing our results of  \cite{susyrevelation} on  soft supersymmetry breaking terms and open-string moduli masses, we  discuss the one-loop RG running of the squark and slepton masses in mSUGRA-like models (using the running of the gaugino masses) to the EW scale in the large volume limit. Phenomenological constraints and some of the calculated soft SUSY parameters identify the $D7$-brane Wilson line moduli as the first two generations/families of squarks and sleptons and the $D3$-brane
  (restricted to the big divisor) position moduli as the two Higgses for MSSM-like models at $TeV$ scale. We also discuss how the obtained open-string/matter moduli make it easier to impose FCNC constraints, as well as RG flow of off-diagonal squark mass(-squared) matrix elements.

\end{abstract}
\end{titlepage}


\section{Introduction}

From the point of view of ``testing" string theory in the laboratories, string phenomenology and string cosmology
 have been the major areas of work.  In  the context of Type IIB orientifold compactification in the L(arge) V(olume) S(cenarios) limit, a non-supersymmetric $AdS$ minimum was realized in \cite{Balaetal2} with the inclusion of perturbative ${\alpha^{\prime}}^3$ corrections
  to the K\"{a}hler potential, which was then uplifted to $dS$ vacuum. Followed by this, it was shown in \cite{dSetal} that with the inclusion of
  (non-)perturbative $\alpha^{\prime}$ corrections to the K\"{a}hler potential and instanton corrections to the superpotential,
  one can realize {\it non}-supersymmetric metastable $dS$ solution in a more natural way without having to add an uplifting term (e.g. with the inclusion of $\overline D3$-brane as in \cite{KKLT}).
Now, type IIB LVS orientifold compactifications with  D7-branes wrapping the smaller cycle
produce qualitatively similar gauge couplings as those of the Standard Model
\cite{conloncal}.
Although LVS models in type IIB compactifications have been exciting steps in the
    search for  realistic models in both cosmology as well as phenomenology, studies thus far in the same have required a hierarchy of compactification volumes ${\cal V}$, requiring $\cal{V}$ $\sim 10^6$ for LVS cosmology (e.g. see
    \cite{kaehlerinflation}) and $\cal{V}$ $\sim 10^{14}$ for LVS phenomenology (e.g. see first two references in \cite{conloncal}). In \cite{susyrevelation}, we had given a proposal for obtaining a $10^{12}GeV$ gravitino in the inflationary era and a $TeV$ gravitino in the present epoch (as a resolution of the aforementioned tension between LVS cosmology and LVS phenomenology) in the same string theoretic setup. Further, the breaking of supersymmetry is encoded in the ``soft terms" and is communicated to the visible sector
  (MS)SM via different mediation processes (eg. gravity mediation, anomaly mediation, gauge mediation). In \cite{susyrevelation}, we had also obtained values of soft supersymmetry breaking parameters including  a (near) universality in the masses, $\hat{\mu}$-parameters, Yukawa couplings  and the $\hat{\mu}B$-terms for the $D3$-brane position moduli - the Higgs doublet in our construction. Building on the same, in this note, we discuss the RG flow to the EW scale of the squark/slepton masses in mSUGRA-like models and use the same to discuss FCNC constraints as well as the RG flow of the squark/slepton off-diagonal mass-squared matrix elements.

In this note, in section {\bf II}, after discussing the basic setup and the method of obtaining ${\cal O}(1)$ $g_{YM}$,  we review our proposal of \cite{susyrevelation} for the resolution of the tension between LVS Cosmology and LVS Phenomenology pertaining to gravitino masses followed by a brief summary  of soft supersymmetry breaking parameters at an intermediate scale which is about a tenth of the GUT scale.  Section {\bf III} contains new results (relative to \cite{susyrevelation}) related to (a) one-loop RG running of  squark and slepton masses in mSUGRA-like models to the EW scale in the large volume limit, (b) identification of open string or matter moduli with (MSSM) Higgses and squarks and sleptons, (c) FCNC constraints and (d) RG-flow of the squark/slepton off-diagonal mass-squared matrix elements. Section {\bf IV} has the conclusions.

\section{The Setup and What Has Been Achieved Thus Far Using It}

In this section, we first describe our setup: type IIB compactification on the orientifold of a ``Swiss-Cheese Calabi-Yau" in the large volume limit including perturbative $\alpha^\prime$-corrections as well as its modular completion, and the instanton-generated superpotential written out respecting the (subgroup, under orientifolding, of) $SL(2,{\bf Z})$ symmetry of the underlying parent type IIB theory, in the presence of a mobile space-time filling $D3-$brane and stacks of $D7$-branes wrapping the ``big" divisor along with magnetic fluxes. This is followed by a summary of what one is able to achieve in the context of string phenomenology using this setup - resolution of the tension mentioned in section {\bf I} between large volume phenomenology and cosmology, as well as evaluation of soft supersymmetry breaking parameters consistent with several checks.

In  \cite{dSetal,largefNL_r_axionicswisscheese}, we  addressed some cosmological issues like $dS$ realization, embedding inflationary scenarios and realizing non-trivial non-Gaussianities in the context of type IIB Swiss-Cheese Calabi Yau orientifold in LVS. This has been done with the inclusion of (non-)perturbative $\alpha^{\prime}$-corrections to the K\"{a}hler potential and non-perturbative instanton contribution to the superpotential. The Swiss-Cheese Calabi Yau we are using, is a projective variety in ${\bf WCP}^4[1,1,1,6,9]$ given as
\begin{equation}
\label{eq:hyper}
x_1^{18} + x_2^{18} + x_3^{18} + x_4^3 + x_5^2 - 18\psi \prod_{i=1}^5x_i - 3\phi x_1^6x_2^6x_3^6 = 0,
\end{equation}
 which has two (big and small) divisors $\Sigma_B(x_5=0)$ and $\Sigma_S(x_4=0)$ . Now in this setup, for studying phenomenological issues, we also include  a single mobile space-time filling $D3$-brane  and stacks of $D7$-branes wrapping the ``big" divisor $\Sigma_B$ as well as supporting $D7$-brane fluxes. In order to have $ g_{YM}\sim O(1)$ for supporting (MS)SM, in the previously studied LVS models with fluxed $D7$-branes, it was argued that $D7$-brane had to wrap the ``small" divisor $\Sigma_S$. Unlike this, in our setup we show the same to be possible with ``big" divisor $\Sigma_B$ due to the possible competing contributions coming from the Wilson line moduli.
Further, we  considered the rigid limit of wrapping of the $D7$-brane around $\Sigma_B$ (to ensure that there is no obstruction to a chiral matter spectrum) which is effected by considering zero sections of $N\Sigma_B$ and hence there is no superpotential generated due to the fluxes on the world volume of the $D7$-brane \cite{jockersetal}.  Working in the $x_2=1$-coordinate patch, for definiteness, and defining $z_1={x_1}/{x_2},\ z_2={x_3}/{x_2},\ z_3=
{x_4}/{x_2^6}$ and $z_4={x_5}/{x_2^9}$ ($z_4$ for later use) three of which get identified with the mobile $D3$-brane position moduli, the non-perturbative superpotential due to gaugino condensation on a stack of $N$ $D7$-branes wrapping $\Sigma_B$ will be proportional to $\left(1+z_1^{18}+z_2^{18}+z_3^3-3\phi_0z_1^6z_2^6\right)^{\frac{1}{N}}$ (See \cite{Ganor1_2,Maldaetal_Wnp_pref}) , which according to \cite{Ganor1_2}, vanishes as the mobile $D3$-brane touches the wrapped $D7$-brane. To effect this simplification, we will henceforth be restricting the mobile $D3$-brane to the ``big" divisor $\Sigma_B$. It is for this reason that we are justified in considering a single  wrapped $D7$-brane, which anyway can not effect gaugino condensation. Further, as in KKLT scenarios as well as in race-track scenarios with large divisor volumes (See \cite{racetrack_small_Wcs} for the latter), it is possible to tune the complex-structure moduli-dependent superpotential to be negligible as compared to the non-perturbative superpotential; we do the same.

The main idea in realizing $O(1)$ gauge coupling in \cite{susyrevelation} is the competing contribution to the gauge kinetic function (and hence to the gauge coupling) coming from the $D7$-brane Wilson line moduli as compared to the volume of the big divisor $\Sigma_B$, after constructing local (i.e. localized around the location of the mobile $D3$-brane in the Calabi-Yau) appropriate involutively-odd harmonic one-form on the big divisor that lies in $coker\left(H^{(0,1)}_{{\bar\partial},-}(CY_3)\stackrel{i^*}{\rightarrow}
H^{(0,1)}_{{\bar\partial},-}(\Sigma_B)\right)$, the immersion map $i$ being defined as:
$i:\Sigma^B\hookrightarrow CY_3$.  This way the gauge couplings corresponding to the gauge theories living on stacks of $D7$ branes wrapping the ``big" divisor $\Sigma_B$ (with different $U(1)$ fluxes on the two-cycles inherited from $\Sigma_B$) will be given by:
$g_{YM}^{-2}=Re(T_B)\sim \mu_3{\cal V}^{\frac{1}{18}}$, $T_B$ being the appropriate ${\cal N}=1$ K\"{a}hler coordinate (See \cite{susyrevelation} and \cite{jockersetal}) and $\mu_3$(related to the $D3$-brane tension)$=\pi/\kappa^2\sim\pi/(\alpha^\prime)^2$,
implying a finite (${\cal O}(1)$) $g_{YM}$ for ${\cal V}\sim10^6$.


The main idea of our proposal pertaining to the geometric resolution of the tension between LVS cosmology and phenomenology, is the motion of a space-time filling mobile $D3$-brane which dictates the gravitino mass via a holomorphic pre-factor in the superpotential - section of (the appropriate) divisor bundle. The gravitino mass is given by:
$m_{\frac{3}{2}}=e^{\frac{K}{2}}W M_p\sim({W}/{{\cal V}})M_p$ in the LVS limit. With appropriate choice of fluxes, the
superpotential in the presence of an $ED3-$instanton is of the type (See \cite{Grimm,Maldaetal_Wnp_pref,Ganor1_2}) :
\begin{eqnarray}
\label{eq:W_np}
& & \hskip -0.5in W \sim W_{np}
\sim \left(1 + z_1^{18} + z_2^{18} + z_3^2 - 3\phi_0z_1^6z_2^6\right)^{n_s}e^{in^sT_s}\Theta_{n^s}({\cal G}^a,\tau)
\sim\frac{\left(1 + z_1^{18} + z_2^{18} + z_3^2 - 3\phi_0z_1^6z_2^6\right)^{n^s}}{{\cal V}^{n^s}},
\nonumber\\
& &
\end{eqnarray}
 where $n^s$ is the $D3$-instanton number and the holomorphic Jacobi theta function of index $n^s$,  $\Theta_{n^s}(\tau,{\cal G}^a)$ (which encodes the contribution of $D1$-instantons in an $SL(2,{\bf Z})$-covariant form), is defined via $\Theta_{n^s}(\tau,{\cal G}^a)=\sum_{m_a}e^{\frac{i\tau m^2}{2}}e^{in^s {\cal G}^am_a}$ where ${\cal G}^a=c^a-
\tau{\cal B}^a, {\cal B}^a\equiv b^a - lf^a$,  $f^a$ being the components of  two-form fluxes valued in $i^*\left(H^2_-(CY_3)\right)$; ${\cal V}^{n^s}$ comes from the (complexified) $\Sigma_S$-divisor volume which is contained in the ${\cal N}=1$ K\"{a}hler coordinate $T_s$ appropriate to our setup (consisting of a mobile $D3$-brane and a stack of $D7$-branes). Note that $\left(1 + z_1^{18} + z_2^{18} + z_3^2 - 3\phi_0z_1^6z_2^6\right)^{n_s}$ represents a one-loop determinant
of fluctuations around the $ED3$-instanton; $e^{in^sT_s}=e^{-n^s{\rm vol}(\Sigma_S)+i...}$ being a section of the inverse divisor bundle $n^s[-\Sigma_S]$, the holomorphic prefactor $(1 + z_1^{18} + z_2^{18} + z_3^2 - 3\phi_0z_1^6z_2^6)^{n_s}$ has to be a section of $n^s[\Sigma_S]$ to compensate and the holomorphic prefactor, a section of $n^s[\Sigma_S]$ having no poles, must have zeros of order $n^s$ on a manifold homotopic to $\Sigma_S$  (See \cite{Ganor1_2}). In (\ref{eq:W_np}), as mentioned earlier on in this section, we
assume, as in KKLT scenarios, a complex-structure moduli-dependent superpotential  much smaller than the non-perturbative superpotential; appropriate sample values of the complex structure moduli which permit very small values of the (Gukov-Vafa-Witten) complex-structure moduli-dependent superpotential relevant to (\ref{eq:hyper}), are obtained in \cite{DDF}, and race-track models with large divisor volumes ((\ref{eq:hyper}) is an example of the same) with complex-structure moduli-dependent superpotential much smaller than the instanton-generated superpotential, are studied in \cite{racetrack_small_Wcs}. Note, even though the hierarchy in the divisor volumes (namely vol$(\Sigma_S)\sim ln {\cal V}$ and vol$(\Sigma_B)\sim {\cal V}^{\frac{2}{3}}$) was obtained in \cite{Balaetal2} in a setup without a mobile space-time filling $D3$-brane and $D7$-branes wrapping a divisor and assuming that the superpotential receives an ${\cal O}(1)$ dominant contribution from the complex structure superpotential, one can argue that using the correct choice of K\"{a}hler coordinates $T_S$ and $T_B$, the divisor volumes continue to stabilize at the same values in our setup wherein the superpotential receives the dominant contribution from the non-perturbative instanton-generated superpotential \footnote{A quick way to see this is to extremize the potential with respect to $\tau_S\equiv{\rm vol}(\Sigma_S)$ and see at what value the same stabilizes assuming in a self-consistent manner that $\tau_B\equiv{\rm vol}(\Sigma_B)$ has been stabilized at ${\cal V}^{\frac{2}{3}}$. Referring to the discussion in \cite{susyrevelation} on the complete K\"{a}hler potential using the $T_B$ and $T_S$ for $D3-D7$ system and the discussion on page 4 in the paragraph preceding the one containing equation (\ref{eq:W_np}), one sees that with the $D3$-brane restricted to $\Sigma_B$, $K=-2 ln[(\mu_3{\cal V}^{\frac{1}{18}}+...)^{\frac{3}{2}} - (\tau_S+\mu_3{\cal V}^{\frac{1}{18}}+...)^{\frac{3}{2}}+\sum_{\beta\in H_2^-}\sum_{m,n\neq(0,0)}n^0_\beta cos(nk.b+mk.c)(...)+...] + K_{cs}$ - the dots denoting the sub-dominant terms and $K_{cs}$ denoting the complex structure K\"{a}hler potential - where $n^0_\beta$ are the genus-zero Gopakumar-Vafa invariants. Further, $W\sim W_{np}\sim e^{-n^s\tau_S}$. One hence sees that $V\sim e^K K^{\tau_S \tau_S}|D_{\tau_S}W_{np}|^2\sim$
$\frac{e^{-2n^s\tau_S}\sqrt{\tau_S + \mu_3{\cal V}^{\frac{1}{18}}}}{\Xi^2}[(\tau_S + \mu_3{\cal V}^{\frac{1}{18}})^{\frac{3}{2}}+\Xi]$$[-n^s+{\sqrt{\tau_S + \mu_3{\cal V}^{\frac{1}{18}}}}/{\Xi}]^2$
where $\Xi\equiv \sum_{\beta\in H_2^-}\sum_{m,n\neq(0,0)}n^0_\beta cos(nk.b+mk.c)(...)$. According to the Castelnuovo's theory of moduli spaces applied to compact projective varieties - see \cite{Klemm_GV} - $n^0_\beta$ are very large. Also, taking ${\cal V}\sim 10^6$, one can approximate ${\cal V}^{\pm\frac{1}{36}}\approx {\cal O}(1)$. One can then show that $V^\prime(\tau_S)=0$ for
$\tau_S\sim{\cal V}^{\frac{1}{18}}\sim ln {\cal V}, n^s\sim{\cal O}(1)$ having already stabilized $\tau_B$ at ${\cal V}^{\frac{2}{3}}$. One can obtain the same result by a more careful calculation without assuming, to begin with, that the large divisor volume has been stabilized at ${\cal V}^{\frac{2}{3}}$.}.

Denoting the extremum position moduli of the mobile $D3$-brane by $z_{i,(0)}$, consider fluctuations about the same given by $\delta z_{i,(0)}$. Defining $P(\{z_{i,(0)}\})\equiv 1 + z_{1,(0)}^{18} + z_{2,(0)}^{18} + z_{3,(0)}^2 - 3\phi_0 z_{1,(0)}^6z_{2,(0)}^6$, one obtains:
$W \sim {\cal V}^{\alpha n^s - n^s}\left(1 + \frac{\sum_i a_i\delta z_{i,(0)}}{P(\{z_{i,(0)}\})}\right)^{n^s},
$ where $a_i$'s are some order one factors and one assumes $P(\{z_{i,(0)}\})\sim {\cal V}^\alpha$. This yields $m_{\frac{3}{2}}\equiv e^{\frac{\hat{K}}{2}}|\hat{W}|\sim{\cal V}^{n^s(\alpha - 1) - 1}$
in the LVS limit. In order to be able to obtain a $10^{12} GeV$ gravitino at ${\cal V}\sim 10^6$ in the inflationary epoch, one hence requires: $\alpha = 1$ ($n^s\geq2$ to ensure a metastable dS minimum in the LVS limit - see \cite{dSetal}) and geometrically, $(z_1,z_2,z_3)\sim({\cal V}^{\frac{1}{18}},{\cal V}^{\frac{1}{18}},z_3)$ along the non-singular elliptic curve:
$\psi_0{\cal V}^{\frac{1}{9}}z_3z_4 - z_3^2 - z_4^3 \sim {\cal V}$.
A similar analysis for obtaining a $TeV$ gravitino in the present epoch would require: $\alpha=1-{3}/{2n^s}$, which for $n^s=2$  yields $\alpha={1}/{4}$, and geometrically, $(z_1,z_2,z_3)\sim({\cal V}^{\frac{1}{72}},{\cal V}^{\frac{1}{72}},z_3)$ along the non-singular elliptic curve:
$\psi_0{\cal V}^{\frac{1}{36}}z_3z_4 - (z_3^2 + z_4^3)\sim{\cal V}^{\frac{1}{4}},$ embedded inside the Swiss-Cheese Calabi-Yau.  Taking the small divisor's volume modulus and the Calabi-Yau volume modulus as independent variables, one can show that the volume of the Calabi-Yau can be extremized at one value - $10^6$ - for varying positions of the mobile $D3$-brane (See \cite{susyrevelation}). Apart from details given in \cite{susyrevelation}, some of the major contributory reasons include (a) the $D3$-brane position moduli enter the holomorphic prefactor in the superpotential and hence the extremization of the overall potential, proportional to the modulus squared of the same, is not influenced by this prefactor, and (b) in consistently taking the large volume limit as done in this paper, the superpotential is independent of the Calabi-Yau volume modulus.

We now summarize our results of \cite{susyrevelation} pertaining to gaugino masses and soft supersymmetry breaking parameters.
The soft supersymmetry parameters are related to the expansion of the K\"{a}hler potential and Superpotential for the open- and closed-string moduli as a power series in the open-string (the ``matter fields") moduli. In our setup, the matter fields - the mobile space-time filling $D3$-brane position moduli in the Calabi-Yau  and the complexified Wilson line moduli arising due to the wrapping of $D7$-brane(s) around four-cycles - take values (at the extremum of the potential) respectively of order ${\cal V}^{\frac{1}{36}}$ and ${\cal V}^{-\frac{1}{4}}$ (verifiable by extremization of the potential), which are finite. The superpotential can be expanded about these values as:
\begin{eqnarray}
\label{eq:W_exp}
& & W\sim{\cal V}^{\frac{n^s}{2}}\Theta_{n^s}(\tau,{\cal G}^a)e^{in^sT(\sigma^S,{\bar\sigma^S};{\cal G}^a,{\bar{\cal G}^a};\tau,{\bar\tau})}[1 + (\delta {\cal Z}_1 + \delta {\cal Z}_2)\{n^s{\cal V}^{-\frac{1}{36}} + (in^s\mu_3)^3{\cal V}^{\frac{1}{36}}\}\nonumber\\
& & +\delta\tilde{{\cal A}}_1\{-[\lambda_1+\lambda_2](in^s\mu_3){\cal V}^{-\frac{31}{36}} - n^s[\lambda_1+\lambda_2]{\cal V}^{-\frac{11}{12}}\}]+ ((\delta {\cal Z}_1)^2 + (\delta {\cal Z}_2)^2)\mu_{{\cal Z}_i{\cal Z}_i}  + \delta {\cal Z}_1\delta {\cal Z}_2\mu_{{\cal Z}_1{\cal Z}_2}\nonumber\\
& & \hskip-1in+
(\delta\tilde{{\cal A}}_1)^2\mu_{\tilde{\cal A}_I\tilde{\cal A}_I} + \delta {\cal Z}_1\delta\tilde{{\cal A}}_1\mu_{{\cal Z}_1\tilde{\cal A}_I} + \delta {\cal Z}_2\delta\tilde{{\cal A}}_1\mu_{{\cal Z}_2\tilde{\cal A}_I} + ((\delta {\cal Z}_1)^3 + (\delta {\cal Z}_2)^3)Y_{{\cal Z}_i{\cal Z}_i{\cal Z}_i} + ((\delta {\cal Z}_1)^2\delta {\cal Z}_2 + (\delta {\cal Z}_2)^2\delta {\cal Z}_1)Y_{{\cal Z}_i{\cal Z}_i{\cal Z}_j}\nonumber\\
& & + (\delta {\cal Z}_1)^2\delta{\tilde{\cal A}}_1Y_{{\cal Z}_i{\cal Z}_i\tilde{\cal A}_I} + \delta {\cal Z}_1(\delta\tilde{\cal A}_I)^2Y_{{\cal Z}_i\tilde{\cal A}_I\tilde{\cal A}_I} + \delta {\cal Z}_1\delta {\cal Z}_2\delta\tilde{{\cal A}}_1Y_{{\cal Z}_1{\cal Z}_2\tilde{\cal A}_I} + (\delta\tilde{\cal A}_I)^3Y_{\tilde{\cal A}_I\tilde{\cal A}_I\tilde{\cal A}_I}+
 ....,
\end{eqnarray}
where $\sigma^S$ (and for later uses, $\sigma^B$) is the volume of the small (big) divisor complexified by the RR four-form axion, the constants $\lambda_{1,2}$ are functions of the extremum values of the closed string moduli and the Fayet-Illiopoulos parameters corresponding to the NLSM (the IR limit of the GLSM) for an underlying ${\cal N}=2$ supersymmetric gauge theory whose target space is the (toric) projective variety we are considering in our work, and the mobile $D3$-brane position moduli fluctuations $\delta{\cal Z}_i$'s and the $D7-$brane Wilson line moduli fluctuations $\delta\tilde{\cal A}_I$'s ($I$ indexes $dim\left(H^{(0,1)}_{{\bar\partial},-}(CY_3)\right)$ and for simplicity we take $I=1$ corresponding to the harmonic one-form referred to earlier on in this section and explicitly constructed in \cite{susyrevelation}), obtained in \cite{susyrevelation}, diagonalize the open string moduli metric.

With a (partial) cancelation between the volume of the ``big" divisor and the Wilson line
contribution (required for realizing $\sim O(1) g_{YM}$ in our setup), in \cite{susyrevelation}, we calculated in the large volume limit: (a) the gravitino and the
gaugino masses of ${\cal O}(1-10^3)TeV$ (for ${\cal V}\sim10^6$ and appropriate values of the $D3$-instanton number $n^s$), (b) the $D3-$brane position moduli masses to be $\sim {\cal V}^{\frac{19}{36}}m_{{3}/{2}}$ and
Wilson line moduli masses to be  $ \sim {\cal V}^{\frac{73}{72}}m_{{3}/{2}}$,
(c) the $\mu$ (defined in (\ref{eq:W_exp})) and the
physical $\hat{\mu}$ parameters defined via:
$\hat{\mu}_{ij}={(\frac{{\bar{\hat{W}}}e^{\frac{\hat{K}}{2}}}{|\hat{W}|}\mu_{ij} + m_{\frac{3}{2}}Z_{ij}\delta_{ij} - {\bar F}^{\bar m}{\bar\partial}_{\bar m}Z_{ij}\delta_{ij})}/{\sqrt{\hat{K}_{i{\bar i}}\hat{K}_{j{\bar j}}}},$
(d) the Yukawa couplings $Y_{ijk}$ (defined in
(\ref{eq:W_exp})), the physical Yukawa couplings $\hat{Y}_{ijk}$ defined via:
$\hat{Y}_{ijk}={e^{\frac{\hat{K}}{2}}Y_{ijk}}/{\sqrt{\hat{K}_{i{\bar i}}\hat{K}_{j{\bar j}}\hat{K}_{k{\bar k}}}},$
the $A_{ijk}$-terms defined via
$A_{ijk}=[\hat{K}_m + \partial_m ln Y_{ijk} - \partial_m ln(\hat{K}_{i{\bar i}}\hat{K}_{j{\bar j}}\hat{K}_{k{\bar k}})]$
 and
 (e) the $\hat{\mu}B$-parameters defined through an involved expression (See \cite{conloncal}) involving $F^m=e^{\frac{\hat{K}}{2}}\hat{K}^{m{\bar n}}{\bar D}_{\bar n}{\bar W}$, $\hat{K}_{i{\bar j}}\equiv\frac{\partial^2 K \left(\left\{\sigma^b,{\bar\sigma}^B;\sigma^S,{\bar\sigma}^S;{\cal G}^a,{\bar{\cal G}}^a;\tau,{\bar\tau}\right\};\left\{\delta z_{1,2},{\bar\delta}{\bar z}_{1,2};\delta{\cal A}_I,{\bar\delta}{\bar{\cal A}_I}\right\}\right)}{\partial C^i{\bar\partial} {\bar C}^{\bar j}}\bigg|_{C^i=0}$ and $Z_{ij}\equiv\frac{\partial^2 K \left(\left\{\sigma^b,{\bar\sigma}^B;\sigma^S,{\bar\sigma}^S;{\cal G}^a,{\bar{\cal G}}^a;\tau,{\bar\tau}\right\};\left\{\delta z_{1,2},{\bar\delta}{\bar z}_{1,2};\delta{\cal A}_I,{\bar\delta}{\bar{\cal A}_I}\right\}\right)}{\partial C^i{\partial} C^{j}}\bigg|_{C^i=0}$ - the matter field fluctuations denoted by $C^i\equiv \delta {\cal Z}_{1,2},\delta{\cal A}_I$, and
$K \left(\left\{\sigma^b,{\bar\sigma}^B;\sigma^S,{\bar\sigma}^S;{\cal G}^a,{\bar{\cal G}}^a;\tau,{\bar\tau}\right\};\left\{\delta z_{1,2},{\bar\delta}{\bar z}_{1,2};\delta{\cal A}_I,{\bar\delta}{\bar{\cal A}_I}\right\}\right)$ denoting the complete K\"{a}hler potential.

Further, the complete K\"{a}hler potential will consist of the following contribution
 $- 2 ln[a(T_B + {\bar T}_B - \gamma K_{\rm geom})^{\frac{3}{2}}
 -a(T_S + {\bar T}_S - \gamma K_{\rm geom})^{\frac{3}{2}} + ...]$ where $\gamma=\kappa_4^2 T_3$, $T_3$ being the $D3-$brane tension, and hence the same requires evaluation of the geometric K\"{a}hler potential $K_{\rm geom}$. Using GLSM techniques and the toric data for (\ref{eq:hyper}), the geometric K\"{a}hler potential for the divisor ${\Sigma_B}$ (and ${\Sigma_S}$) in the LVS limit was evaluated in \cite{susyrevelation} in terms of derivatives of genus-two Siegel theta functions as well as two Fayet-Iliopoulos parameters corresponding to the two $C^*$ actions in the  two-dimensional ${\cal N}=2$ supersymmetric gauge theory whose target space is our toric variety Calabi-Yau, and a parameter $\zeta$ encoding the information about the $D3-$brane position moduli-independent (in the LVS limit) period matrix of the hyperelliptic curve $w^2=P_{\Sigma_B}(z)$, $P_{\Sigma_B}(z)$ being the defining hypersurface for $\Sigma_B$. The geometric K\"{a}hler potential scales with the Swiss-Cheese Calabi-Yau's volume as ${{\cal V}^{\frac{2}{3}}}/{\sqrt{ln {\cal V}}}$.

 In addition, we had found that ${\hat{\mu}}^2 \sim {\hat{\mu} B}$ for the $D3$-brane position moduli to be consistent with the requirement of a stable vacuum spontaneously breaking supersymmetry - see \cite{Green_Weigand} - whereas ${\hat{\mu}}^2 \ll {\hat{\mu} B}$ for components with only Wilson line modulus as well as the same  mixed with the $D3$-brane position moduli. Also, the un-normalized physical mu-parameters for the $D3$-brane position moduli  are $\sim$ TeV, as required for having correct electroweak symmetry breaking in the context of ``$\mu$-problem"\cite{Green_Weigand,mu1}.

\section{RG Flow of Squark and Slepton Masses to the EW Scale}
In the context of string phenomenology, the study of the origin and dynamics of SUSY breaking are among the most challenging issues and several proposals for the origin of SUSY breaking as well its transmission to the visible sector with a particular structure of soft parameters have been studied. For addressing realistic model-building issues, the gaugino masses as well as other soft SUSY-breaking parameters have to be estimated at low energy which requires the study of their running to electroweak scale using the respective RG-equations, imposing the low energy FCNC constraints. Further the ratio of gaugino masses to the square of gauge couplings ($M_a/{g_a}^2$), are well-known RG-invariants at one loop  as their RG-running up to two-loops are given as \cite{sparticlesreview}:
\begin{eqnarray*}
& & {dg_a}/{dt} = {{g_a}^3b^a}/{16 {\pi}^2}  + ({{g_a}^3}/{16 {\pi}^2})[\sum^3_{b=1}B^{(2)}_{ab}{g_b}^2- {{1}/{16 \pi^2}}\sum_{x=u,d,e,\nu}{C^a_x}/{16 \pi^2} Tr[{Y_x}^{\dagger}{Y_x}]] \nonumber\\
& & \hskip-0.9in{dM_a}/{dt} = {2{g_a}^2b^a M_a}/{16 {\pi}^2}  + ({2{g_a}^2}/{({16 {\pi}^2})^2})[\sum^3_{b=1}B^{(2)}_{ab}{g_b}^2 (M_a+M_b)+ \sum_{x=u,d,e,\nu} {C^a_x}\{Tr[{Y_x}^{\dagger}{\tilde{A_x}}]-M_a Tr[{Y_x}^{\dagger}{Y_x}]\}]\nonumber\\
\end{eqnarray*}
\noindent where $t=ln({{Q_{EW}}}/{Q_0})$ defined in terms of ${Q_{EW}}$ which is the phenomenological low energy scale (of interest) and $Q_0$ some high energy scale alongwith the MSSM gauge coupling $\beta$-functions' given as $b^a=\{{{33}/{5},1,-3}\}$, $B^{(2)}_{ab}$ and $C^a_x$ being $3\times3$ and $4\times3$ matrices with ${\cal O}(1-10)$ components and ${\tilde{A_x}}$, $Y_x$ are trilinear $A$-term and Yukawa-coupling respectively. Further, the first term on the right hand sides of each of above equations represents one-loop effect while other terms in the square brackets are two-loop contributions to their RG running implying that $ d/dt[{M_a}/{{g_a}^2}]=0$ at one-loop.

RG equations of first family of squark and slepton masses result in the following set of equations  which represent the difference in their mass-squared values between  ${Q_{EW}}$ and  $Q_0$ at one-loop level {\cite{Martinreview,QuevedoLHC}}:
\begin{eqnarray}
\label{eq:RGsparticles}
& & {M^2_{\tilde{d}_L,\tilde{u}_L}}|_{Q_{EW}}-M^2_{\tilde{d}_L,\tilde{u}_L}|_{Q_0}={\cal K}_3+{\cal K}_2+{\frac{1}{36}}{\cal K}_1+ \tilde{\Delta}_{\tilde{d_L}}\nonumber\\
& & {M^2_{\tilde{d}_R}}|_{Q_{EW}}-M^2_{\tilde{d}_R}|_{Q_0}={\cal K}_3+{\frac{1}{9}}{\cal K}_1+ \tilde{\Delta}_{\tilde{d_R}}\nonumber\\
& & {M^2_{\tilde{u}_R}}|_{Q_{EW}}-M^2_{\tilde{u}_R}|_{Q_0}={\cal K}_3+{\frac{4}{9}}{\cal K}_1+ \tilde{\Delta}_{\tilde{u_R}}\nonumber\\
& & {M^2_{\tilde{e}_L}}|_{Q_{EW}}-M^2_{\tilde{e}_L}|_{Q_0}={\cal K}_2+{\frac{1}{4}}{\cal K}_1+ \tilde{\Delta}_{\tilde{e_L}}\nonumber\\
& & {M^2_{\tilde{e}_R}}|_{Q_{EW}}-M^2_{\tilde{e}_R}|_{Q_0}={\cal K}_1+ \tilde{\Delta}_{\tilde{e_R}}\nonumber\\
& & {M^2_{\tilde{\nu}}}|_{Q_{EW}}-M^2_{\tilde{\nu}}|_{Q_0}={\cal K}_2+{\frac{1}{36}}{\cal K}_1+ \tilde{\Delta}_{\tilde{\nu}}
\end{eqnarray}
where the parameters ${\cal K}_a$ are defined through integral (\ref{eq:Kintegrals}) below and the difference in the coefficients of ${\cal K}_1$ in the above set of solutions to respective RG equations is due to various weak hypercharge-squared values for each scalar:
\begin{equation}
\label{eq:Kintegrals}
{\cal K}_a\sim {{\cal O}({{1}/{10}})}\int_{ln Q_0}^{ln Q_{\rm EW}} dt g_a^2(t)M_a^2(t) \equiv {{\cal O}(1/10)} ({M_a}/{g_a^2})^2|_{Q_0}[g_a^4|_{Q_{EW}}-g_a^4|_{Q_0}]_{\rm 1-loop}.
\end{equation}
Further ${\tilde{\Delta}_{\tilde{x}}}$ (appearing in (\ref{eq:RGsparticles})), where ${\tilde{x}} \in \{{\tilde{d_L}},{\tilde{d_R}},{\tilde{u_L}},{\tilde{u_R}},{\tilde{e_L}},{\tilde{e_R}},{\tilde{\nu}}\}$ (i.e. the first family of squarks and sleptons)  are D-term contributions which are ``hyperfine" splitting in squark and slepton masses arising due to quartic interactions among squarks and sleptons with Higgs. These ${\tilde{\Delta}_{\tilde{x}}}$ contribution are generated via the neutral Higgs acquiring VEVs in electroweak symmetry breaking and are of the form \cite{Martinreview}:
$
{\tilde{\Delta}_{\tilde{x}}}\equiv[T_{3{\tilde{x}}}-Q_{{\tilde{x}}} {\rm Sin}^2(\theta_W)]{\rm Cos}(2\beta) m^2_Z$,
where $T_{3{\tilde{x}}}$ and $Q_{{\tilde{x}}}$ are third component of weak isospin and the electric charge of the respective left-handed chiral supermultiplet to which ${\tilde{x}}$ belong. The angle $\theta_W$ is electroweak mixing angle, $m_Z\sim100{\rm GeV}$ and  ${\rm tan}(\beta)$ is the ratio of vevs of the two Higgs  after electroweak symmetry breaking. Now, in our setup $Q_0\equiv{M_{\rm string}}={M_{\rm GUT}}/{10}\sim 10^{15}GeV$ and $Q_{\rm EW}\sim TeV$. As argued in \cite{sparticlesreview,Martinreview,QuevedoLHC}, up to one loop, ${d}/{dt}({M_a}/{g_a^2})=0$. Hence, the aforementioned ${\cal K}_a$-integrals (\ref{eq:Kintegrals}) can be written as:
\begin{equation}
{\cal K}_{a}\sim {{\cal O}(1/10)} ({M_a}/{g_a^2})^2|_{Q_0}[g_a^4|_{Q_{EW}}-g_a^4|_{Q_0}]_{\rm 1-loop}
\end{equation}
As argued in \cite{Kap_Louis}, the gauge couplings run as follows (up to one loop):
\begin{eqnarray}
\label{eq:RG_flow_1}
{16\pi^2}/{g_a^2(Q_{EW})}={16\pi^2}/{g_a^2(Q_0)} + 2b_a ln[{Q_0}/{m_{{3}/{2}}}] + 2b_a^\prime ln[{m_{{3}/{2}}}/{Q_{EW}}] + \Delta_a^{\rm 1-loop},
\end{eqnarray}
where $b_a, b_a^\prime$ are group-theoretic factors and $\Delta_a^{\rm 1-loop}\sim {\bf Tr} {ln ({{\cal M}}/{m_{{3}/{2}}})},$ and  ${\cal M}\equiv e^K\hat{K}^{-\frac{1}{2}}\mu^\dagger(\hat{K}^{-1})^T\mu\hat{K}^{-\frac{1}{2}}$ (See \cite{Kap_Louis}), in our setup, is to be evaluated for the $D7$ Wilson-line modulus ${\cal A}_1$'s. One can show that ${\cal M}_{{\cal A}_1}\sim{\cal V}^{-\frac{13}{6}}$ and  $m_{{3}/{2}}\sim{\cal V}^{-\frac{n^s}{2}-1}$. Hence, from (\ref{eq:RG_flow_1}) one obtains:
$${16\pi^2}/{g_a^2(Q_{EW})}={16\pi^2}/{g_a^2(Q_0)} + {\cal O}(10).$$
As argued in \cite{susyrevelation}, in anomaly-mediated scenarios, ${M_a}/{g_a^2}|_{Q_0}\sim{{\cal V}^{\frac{1}{18}}m_{{3}/{2}}}/{8\pi^2}$ which, bearing in mind that the same is generically suppressed relative to the gravity-mediation result by about ${1}/{8\pi^2}$ (See \cite{susyrevelation}), implies
${M_a}/{g_a^2}|_{Q_0}\sim{\cal V}^{\frac{1}{18}}m_{{3}/{2}}$ as the gravity-mediation result. Further,
using the result: ${1}/{g_a^2}\sim{\cal V}^{\frac{1}{18}}$ of section {\bf II}, one obtains from (\ref{eq:Kintegrals}) :
\begin{eqnarray*}
& & K_a\sim {{\cal O}\bigg({\frac{1}{10}}\biggr)}{\cal V}^{\frac{1}{9}}m_{{3}/{2}}^2(-{\cal V}^{-\frac{1}{9}}+{(16\pi^2)^2}/{[{\cal O}(10)+16\pi^2{\cal V}^{\frac{1}{18}}]^2})\sim m_{{3}/{2}}^2\{{{\cal O}(1)}/{340}\}
\end{eqnarray*}
for ${\cal V}\sim10^6$. Hence, for $m_{{3}/{2}}\sim10TeV$ (which can be realized in our setup - see \cite{susyrevelation}), one obtains $K_a\sim0.3(TeV)^2$ to be compared with $0.5(TeV)^2$ as obtained in \cite{QuevedoLHC}; an mSUGRA point on the ``SPS1a slope" has a value of around $(TeV)^2$. Further the ${\tilde{\Delta}_{\tilde{x}}}$ contributions, being proportional to $m^2_Z$, are suppressed as compared to ${\cal K}_a$-integrals at one-loop.

Further, as suggested by the phenomenological requirements, the first and second family of squarks and sleptons with given gauge quantum number are supposed to possess (approximate) universality in the soft parameters. However, the third family of squark and sleptons, feeling the effect of larger Yukawa's, can get normalized differently. For our setup, we have $D3$-brane position moduli and $D7$-brane Wilson line moduli, which could be suitable candidates to be identified with the Higgs, squarks and sleptons of the MSSM, given that they fulfil the phenomenological requirements. In our setup, one can see  that at a string scale of $10^{15}GeV$,  $D3$-brane position moduli masses are universal with a value of the order $10^4 \rm TeV$ (as ${m_{{\cal Z}_i}}\sim{\cal V}^{\frac{19}{36}} m_{\frac{3}{2}}\sim 10^4 TeV$ for ${\cal V}\sim 10^6$ corresponding to a $10 \rm TeV$ gravitino)  and may be identified with the two Higgses of MSSM spectrum. The Wilson line moduli have masses $\sim {\cal V}^{\frac{73}{72}} m_{\frac{3}{2}}\sim 10^7 TeV$ and are hence heavier than the $D3$-brane position moduli. Moreover, in our Swiss-Cheese orientifold setup the trilinear $A$-terms show universality and are calculated to be $\sim 10^7 {\rm TeV}$ along with the physical Yukawa-couplings which are found to be in the range from a negligible value $\sim{\cal V}^{-\frac{85}{24}}\sim 10^{-21}$ (for purely Wilson line moduli contributions) to a high value $\sim{\cal V}^{\frac{43}{24}}\sim 10^{11}$ (for purely $D3$-brane position moduli contributions). As suggested by phenomenology, the first and second family of squarks and slepton masses involve negligible Yukawa-couplings. The Wilson line moduli in our setup could hence be identified with the first and second family of squarks and sleptons. Further, within the one-loop results and dilute flux approximation in our Swiss-Cheese LVS setup, gaugino masses are (nearly) universal with $m_{\rm gaugino}\sim m_{\frac{3}{2}}\sim 10{\rm TeV}$ at the string scale  $M_s\sim10^{15}GeV$ which being nearly the GUT scale would imply that the gauge couplings are almost unified.

Next let us elaborate on low energy FCNC-constraints in our setup. The smallness of FCNC imposes tight constraints on the off-diagonal terms of the scalar mass matrices at low energy ($Q_{EW}$). For example, the FCNC-constraints on the Kaon system demands the following \cite{FCNC}:
\begin{eqnarray}
\label{eq:FCNC}
& & \left(\begin{array}{c}Re\\ Im \end{array}\right)[{(M^d_{\rm off})^2_{LL\ {\rm or}\ RR}}/{M^2_{\rm av}}]\le
\left(\begin{array}{c}{10}^{-2}\\ 10^{-3}\end{array}\right) \frac{M_{\rm av}}{Q_{EW}}
\end{eqnarray}
where ${(M^d_{\rm off})_{LL\ {\rm or}\ RR}}$ represent the off-diagonal entry (with respect to family indices) of left-handed or right-handed down sector squark mass-squared matrix in the ``super CKM-basis" \cite{sparticlesreview} respectively and ${M^2_{\rm av}}$ is the average squark mass; both sides of (\ref{eq:FCNC}) are to be evaluated at $Q_{EW}$.  In the LVS models studied thus far, the low energy FCNC constraints of (\ref{eq:FCNC}) imply a tight window for the allowed off-diagonal mass squared elements as the scalar/gaugino masses and the other soft parameters are found to be of the order of the gravitino mass (or suppressed by an $\sim {\cal O}(10)$ logarithmic factor) which is phenomenologically of the order of ${Q_{\rm EW}}\sim{\rm TeV}$ and hence the off diagonal mass squared components are constrained to be suppressed  by at least a factor of $ {\cal O} ({10^{-4}}-{10^{-3}})$ as compared to the diagonal (average mass squared) values. Unlike this, in our setup, we have realized the scalar masses $10^7 {\rm GeV}$ for $D3$-brane position moduli and $10^{10}{\rm GeV}$ for Wilson line moduli at string scale $M_s\sim10^{15}$ GeV. Further we have noticed by the one loop RG evolution estimates that the difference in the mass-squared values at a scale $M_s\sim10^{15}$ GeV and a TeV scale comes out to be negligible ($\sim ({\rm TeV})^2$) as compared to the mass-squared value ($\sim (10^8-10^{14}) {\rm TeV}^2$). As we have argued that one can identify Wilson line moduli (with masses $\sim {\cal V}^{\frac{73}{72}} m_{\frac{3}{2}}\sim 10^{10} GeV$) with squarks in our setup, this results in a large broadening of the allowed window in off-diagonal entries of mass-squared matrices consistent with (\ref{eq:FCNC}).

Further in the approximation of vanishing Yukawa couplings (which is true in the context of Wilson line moduli in our setup), the diagonal and off-diagonal (w.r.t. the three generation/family indices) entries of squark masses squared evolve differently in the presence of non-zero off-diagonal  trilinear $A$-terms in the respective one-loop RG evolutions and the same result in the following set of equations \cite{FCNC}:
\begin{eqnarray}
& & {\big({M^{2(u/d)}_{LL\ {\rm or}\ RR}}\big)_{ij,\hskip 0.05in i\ne{j}}}\bigg|_{{Q_{EW}}}={\big({M^{2(u/d)}}\big)_{ij,\hskip 0.05in i\ne{j}}}
\bigg|_{M_{\rm GUT}}-{\cal O}(1) \biggl({({{A^{2(u/d)}}})_{ij,\hskip 0.05in i\ne{j}}}\biggr)
\end{eqnarray}
where $({{A^{(u/d)}}})_{ij,\hskip 0.05in i\ne{j}}, (i,j=1,2,3$ corresponding to the three generations/families)
are off-diagonal trilinear A-terms. 
Now,  the above equation implies that if the off-diagonal squark masses squared at $M_{\rm string}\sim10^{15}{\rm GeV}$ are comparable to the contributions of the off-diagonal components of trilinear $A$-term matrix element (which are $\sim {{\cal V}^{\frac{37}{36}}}m_{\frac{3}{2}}\sim 10^{10} {\rm GeV}$ for our setup with ${\cal V}\sim 10^6$) squared, there is then the possibility of a cancelation and can result in small value of the off-diagonal squark masses squared at the low energy scale $Q_{EW}$ that would further enable the low energy FCNC-constraints (\ref{eq:FCNC}) to be satisfied.

\section{Summary and Discussion}
In \cite{susyrevelation}, we had shown that within a setup of a mobile $D3-$brane and a stack of $D7$-branes wrapping the big divisor in Swiss-Cheese Calabi-Yau orientifold compactifications, it is possible to (i) obtain a $10^{12}GeV$ gravitino in the inflationary era and a $TeV$ gravitino in the present epoch, (ii) realize $\sim O(1)$ $g_{YM}$ (in fact even after including renormalization and string loop corrections - see \cite{susyrevelation}) with $D7$-branes wrapping the big divisor in the rigid limit implying the new possibility of supporting SM on $D7$-branes wrapping the big divisor, (iii) a (near) universality in the masses, $\hat{\mu}$-parameters, Yukawa couplings  and the $\hat{\mu}B$-terms for the $D3$-brane position moduli - the two Higgses in our construction - and a hierarchy in the same set and a universality in the $A$ terms on inclusion of the $D7$-brane Wilson line moduli.  Building up on the phenomenological aspects of \cite{susyrevelation}, in this note, we discuss the RG flow of the slepton and squark masses to the EW scale and in the process show that related integrals are close to the mSUGRA point on the ``SPS1a slope".  Based on phenomenological intuitions we further argue that the Wilson line moduli can be identified with the squarks (sleptons)   (at least the first and second families) of MSSM as the Yukawa couplings for the same are negligible; the non-universality in the Yukawa's for the Higgses and squarks, is hence desirable. Finally, we elaborate on how the low energy FCNC-constraints can be more easily satisfied given the large squark masses, as well as how (universal) values of the $A$-parameters can be used to obtain small  off-diagonal entries of mass-squared matrix of squarks at low energy scales again relevant to satisfying the aforementioned FCNC constraints.  A detailed numerical analysis for solving the RG evolutions will definitely explore some more interesting phenomenology in the context of reproducing MSSM spectrum in this LVS Swiss-Cheese orientifold setup.

\section*{Acknowledgements}

One of us (PS), is supported by a Senior Research Fellowship from the CSIR, India. AM would like to thank the Harvard theory group, the SNS at IAS, Princeton, the McGill university theory group and the Abdus Salam ICTP (under the junior associateship program) for their kind hospitality and support where part of this work was done, J.Maldacena and C.Vafa for useful discussions and R.Blumenhagen, O.Ganor, T.Grimm and H.Jockers for useful clarifications.


\end{document}